# Vortex-induced Vibrations of a Confined Circular Cylinder for Efficient Flow Power Extraction

**Atul K. Soti[1]†, and Ashoke De[1]**

[1]Department of Aerospace Engineering, IIT Kanpur, Kanpur, Uttar Pradesh, 208016, India

A simple method to increase the flow power extraction efficiency of a circular cylinder, undergoing vortex-induced vibration (VIV), by confining it between two parallel plates is proposed. A two-dimensional numerical study was performed on VIV of a circular cylinder inside a parallel plate channel of height $H$ at Reynolds number 150 to quantify the improvement. The cylinder is elastically mounted with a spring such that it is only free to vibrate in the direction transverse to the channel flow and has a fixed mass ratio ($m^*$) of 10. The energy extraction process is modelled as a damper, with spatially constant damping ration ($\zeta$), attached to the cylinder. The simulations are performed by varying the reduced velocity for a set of fixed mass-damping ($\alpha = m^*\zeta$) values ranging between 0 to 1. The blockage ratio ($b = D/H$) is varied from 0.25 to 0.5 by changing the channel height. With increasing blockage, the vibration amplitude of the cylinder is found to decrease while the lock-in region shifts towards smaller reduced velocities due to increased Strouhal number for the stationary cylinder case. The quasi-periodic initial branch found for the unconfined cylinder shrinks with the increasing blockage. The extracted power is found to increase rapidly with the blockage. For maximum blockage ($b = 0.2$), the maximum flow power extracted by the cylinder is an order of magnitude larger as compared to what it would extract in an open domain with free stream velocity equal to the channel mean velocity. The optimal mass-damping ($\alpha_c$) for extracting maximum power is found to lie between 0.2 to 0.3. An expression is derived to predict the maximum extracted power from the undamped response of a confined/unconfined cylinder. With the assumption $\alpha_c = 0.25$, the derived expression can predict the maximum power extraction within ±20% of the actual values obtained from present and previous numerical and experimental studies.

## Nomenclature

| | |
|---|---|
| $\alpha$ | mass-damping parameter |
| $\alpha_c$ | optimal mass-damping for maximum power extraction |
| $\ddot{y}$ | Transverse acceleration of the cylinder |
| $\Delta t$ | Time step |
| $\Delta x$ | Minimum mesh size |
| $\dot{y}$ | Transverse velocity of the cylinder |
| $\eta$ | Power extraction efficiency |

† Email address for correspondence: atulsoti@gmail.com

| | |
|---|---|
| $\nu$ | Kinematic viscosity of the fluid |
| $\overline{P}$ | Average power |
| $\overline{P}_\alpha$ | peak average power at $\alpha$ mass-damping |
| $\overline{P}_m$ | Maximum average power |
| $\phi_T$ | phase difference between lift force and displacement |
| $\rho$ | Fluid density |
| $\zeta$ | Damping ratio |
| $A^*$ | Non-dimensional vibration amplitude |
| $A_0^*$ | Maximum vibration amplitude |
| $c$ | Damping coefficient for the linear damping |
| $C_{Ls}$ | Lift coefficient for the stationary cylinder |
| $C_L$ | Lift coefficient for the vibrating cylinder |
| $D$ | Cylinder diameter |
| $f$ | Normalized vibration frequency |
| $f_n$ | Natural frequency of the cylinder |
| $f_u$ | vortex shedding frequency |
| $F_y$ | lift force |
| $H$ | channel height |
| $k$ | Stiffness of the spring |
| $m$ | Total moving mass of the cylinder |
| $m^*$ | Mass ratio |
| $m_f$ | Mass of the displaced fluid |
| $P$ | Instantaneous power |
| $Re$ | Reynolds number |
| $U$ | Mean velocity at channel inlet |
| $U^*$ | Reduced velocity |
| $U^*_{c,\alpha}$ | optimal reduced velocity for peak power extraction at $\alpha$ mass-damping |
| $U^*_{c,0}$ | optimal reduced velocity for maximum vibration amplitude at zero mass-damping |
| $U_\infty$ | Free stream velocity |
| $U_{max}$ | center line velocity at channel inlet |
| $Y$ | Non-dimensional transverse displacement of the cylinder |
| $y$ | Transverse displacement of the cylinder |

## 1. Introduction

Vortex-induced vibrations (VIV) of an elastically mounted circular cylinder with the transverse degree of freedom has been extensively studied because of its practical relevance in engineering applications such as marine risers, buildings, bridges, heat exchangers, etc. A detailed description of the underlying mechanism and vibration response of the VIV can be found in Khalak & Williamson (1999); Williamson & Govardhan (2004); Sarpkaya (2004); Gabbai & Benaroya (2005); Bearman (2011); Wu *et al.* (2012). One of the key non-dimensional parameters in VIV is called reduced velocity ($U^*$) which is defined as the ratio of the characteristic velocities of flow and structure, i.e., $U^* = U/(f_n D)$ where $U$, $f_n$ and $D$ are the free stream velocity, natural frequency and diameter of the cylinder. For a certain range of the reduced velocity, the vortex shedding frequency behind the movable cylinder gets locked to the natural frequency of the cylinder, which is called the lock-in phenomenon (Khalak & Williamson 1999). The cylinder vibrates with a significant amplitude in the lock-in region (range of reduced velocity where the lock-in happens). The maximum vibration amplitude at low Reynolds number ($Re < 200$) is close to $0.6D$ while the same can reach up to $1.0D$ at higher $Re$ (Williamson & Govardhan 2004). Govardhan & Williamson (2006), using previous and their experimental results, found a functional relationship between the maximum vibration amplitude and the Reynolds number given by $A_0^* = \log_{10}(0.41 Re^{0.36})$ for $Re = 500 – 33{,}000$.

The traditional way of extracting flow energy has been the use of wind/water turbines. These turbines need a minimum flow speed (called cut-in speed) to generate a useful amount of power and even a larger rated wind speed for maximum power generation (Manwell *et al.* 2010). On the other hand, it is possible to extract a significant portion of flow energy at small flow speeds using VIV by tuning the natural frequency of the system to achieve lock-in. The VIV turbine is simple in design and more environmentally friendly due to a smaller flow blockage allowing easy passage for aquatic animals. In recent years, the potential of VIV for renewable energy has been realized by many researchers. Bernitsas *et al.* (2008) introduced the idea of vortex-induced vibration aquatic clean energy (VIVACE) to extract flow energy using VIV of a cylinder. VIVACE can extract energy from slow ocean currents at greater efficiency as compared to the traditional wave-energy converters. Lee & Bernitsas (2011) conducted high $Re$ and high damping experiments with VIVACE and found a maximum $33\%$ of the total flow energy ($\frac{1}{2}\rho U^3$ per unit area) crossing through the cylinder cross-section ($DL$ where $L$ is the cylinder length) could be harnessed at Reynolds number 75,000. Barrero-Gil *et al.* (2012) used the force measurement data of Hover *et al.* (1998) and Gopalkrishnan (1992) at $Re \sim 10^3$ and $10^4$, respectively, for a forced vibrating cylinder to calculate its VIV response. They concluded that the power extraction capability of the cylinder increased with the Reynolds number. The maximum power extraction was found to be close to $18\%$ and $24\%$ of the total available power through the cylinder cross-section for the two Reynolds number cases. Ding *et al.* (2015) conducted two-dimensional unsteady RANS (Reynolds-Averaged Navier–Stokes) simulations for $10{,}000 < Re < 130{,}000$ to study flow-induced motion (FIM) of circular, square, trapezoid and triangular prisms. The maximum energy extraction of $45.7\%$ and $37.9\%$ were found for circular and trapezoid cross-sections, respectively, at $Re = 60{,}000$. Soti *et al.* (2018) experimentally studied the effects of damping on the VIV response of a circular cylinder and quantified the extracted power. The power extraction was found to be close to $15\%$, $18\%$ and $20\%$ of the total available power through the cylinder cross-section for $Re \sim 1747$, $3107$ and $5328$, respectively. Soti *et al.* (2017), using two-dimensional numerical simulations, found the same to be $10\%$, $13\%$ and $14\%$ for $Re = 100$, $150$ and $200$, respectively.

According to the Betz limit (Betz 1966), a turbine can at most extract $16/27$ ($59.3\%$) fraction of the total flow energy that is crossing through the turbine ($= \frac{1}{2}\rho U^3 S$ where $S$ is the turbine frontal area). In case of VIV, although the cylinder obstructs $DL$ area of the flow at any instant, it covers an area of $(D + 2A)L$ where $A$ is the vibration amplitude. Therefore, the power extraction efficiency of the cylinder is defined based on the total area covered by the vibrating cylinder. The efficiency of the circular cylinder varies from $7.5\%$ to $8.3\%$ for low ($=150$) to moderate ($= 5000$) Reynolds number Soti *et al.* (2017, 2018). The conventional wind turbines are unlikely to work at the flow velocities corresponding to such low Reynolds numbers. It is possible to increase the efficiency of a turbine beyond the Betz limit using confinement (Garrett & Cummins 2007; Gilbert & Foreman 1983). The present work is inspired by the idea of confinement for increasing the flow power extraction efficiency. The effects of confinement on VIV are also not known in the literature. The objective of the present work is to study the effect of confinement on VIV of the circular cylinder and its flow power extraction efficiency. Two-dimensional numerical simulations of flow inside a channel containing an elastically mounted cylinder are conducted. The height of the channel is varied between $2D$ to $4D$ to vary confinement.

The paper is organised as follows. The problem statement is defined in section 2. The flow and structure solvers are described in section 3.1 and 3.2, respectively. A code validation study is presented in section 3.3. The mesh and domain independence studies are presented in section 3.4.

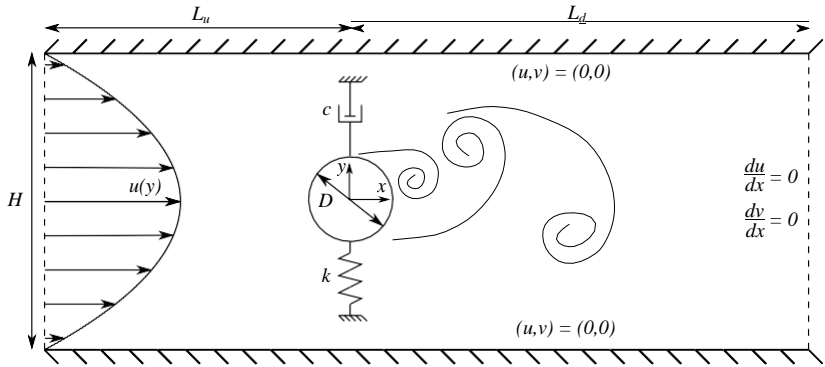

Figure 1: Computational domain and boundary conditions.

VIV of an unconfined cylinder is discussed in section 4.1 for comparing it with the confined case. The effects of confinement on the lift force and Strouhal number for the confined cylinder is presented in section 4.2 which are then used for predicting the effects of confinement on VIV. The vibration response of the undamped cylinder for various blockages is presented in section 4.3. The effect of damping in the vibration response is discussed in section 4.4. The amount of power extracted by the cylinder is presented in section 4.5. Finally, the effect of confined on power extraction efficiency of the cylinder is discussed in section 4.6.

## 2. Problem statement

In the present work, flow power extraction using vortex-induced vibrations (VIV) of a circular cylinder in channel flow is investigated using numerical simulations. The computational domain of the problem is shown in figure 1. A rigid circular cylinder of diameter $D$ is kept inside a parallel plate channel of height $H$. The cylinder is located at a distance of $L_u$ and $L_d$ from inlet and outlet of the channel, respectively. The cylinder is elastically mounted to allow only the transverse motion with the help of springs having the effective spring constant $k$. The cylinder with unstretched springs is located midway along the channel height. The energy extraction process is modelled as a viscous damper with a constant damping coefficient $c$ in parallel to the springs. Soti *et al.* (2017) have shown that the constant damping model is capable of correctly predicting the average power extracted by a more realistic power extraction mechanism such as an electromagnetic generator. The inlet of the channel is specified as a fully developed parabolic velocity profile

$$u(y) = \frac{3U}{2}\left[1 - \frac{4y^2}{H^2}\right], \qquad (2.1)$$

where $U$ represents the mean flow velocity at the inlet. No-slip velocity boundary condition is applied at the channel walls and the cylinder surface. At the channel outlet, zero Neumann boundary condition is used for flow velocity.

## 3. Numerical methodology

### 3.1. Flow solver

The simulation methodology works by numerically integrating the coupled differential equations governing the fluid flow and the cylinder motion. The fluid is assumed to be incompressible and is governed by the following non-dimensional Navier-Stokes equations

$$\frac{\partial \mathbf{u}}{\partial t} + (\mathbf{u} \cdot \nabla)\mathbf{u} = -\nabla p + \frac{1}{Re}\nabla^2 \mathbf{u}, \qquad (3.1)$$

along the with continuity equation

$$\nabla \cdot \mathbf{u} = 0, \qquad (3.2)$$

where $\mathbf{u}$ and $p$ represents the non-dimensional flow velocity vector and fluid kinematic pressure, respectively. Let the kinematic viscosity of the fluid be represented by the symbol $\nu$ and the reference velocity and length scales for nondimensionalization be $U$ and $D$, respectively, then the Reynolds number is defined as $Re = UD/\nu$. In the present case, $U$ represent the mean velocity of the channel flow. The Reynolds number is fixed at $Re = 150$ so that the flow can be assumed two-dimensional.

An in-house fluid-structure interaction (FSI) solver was utilized for performing the numerical simulations of VIV of the circular cylinder in a laminar channel flow. The FSI solver employs a sharp interface immersed-boundary (IB) method based flow solver. The IB method employs a non-conformal Cartesian mesh on which the flow governing equations are discretised using the finite difference method. A three-dimensional ghost cell methodology is used to account for the fluid-structure interface. In this method, the mesh nodes are labelled as either fluid or solid nodes. Fluid/solid nodes are those mesh points which lie inside the fluid/solid domain. The flow equations are only discretised at the fluid nodes. Some fluid nodes will have at least one out of four (eight in 3D) neighbours as a solid node. The flow solution at such fluid nodes will require solution at the neighbouring solid nodes. Such solid nodes are relabelled as the ghost nodes. The solution at the ghost nodes is found by applying boundary conditions (for example no-slip for velocity) at the fluid-structure interface. This is done by drawing a normal to the fluid-structure interface from ghost node which cuts the interface at a point called boundary intercept. The normal is further extended into the fluid region to a point called image point such that the boundary intercept is midway between the image point and the ghost node. The flow solution at a image point can be interpolated from neighbouring fluid nodes since the image point lies inside the fluid domain. The solutions at a image point and boundary intercept are linearly extrapolated to their ghost point. Coming to the discretization of the flow equations, all spatial derivatives are approximated by the second-order accurate finite-difference central difference schemes. The time integration is performed using a second-order accurate Crank-Nicholson scheme. A complete description of the flow solver was provided by Mittal *et al.* (2008).

### *3.2. Structure motion*

The motion of the cylinder is governed by the following non-dimensional equation (Soti *et al.* 2017)

$$\ddot{Y} + \frac{4\pi\zeta}{U^*}\dot{Y} + \frac{4\pi^2}{(U^*)^2}Y = \frac{2\,C_L}{\pi\,m^*}, \qquad (3.3)$$

where $Y = y/D$ and $C_L = Y_y/(\frac{1}{2}\rho U^2 D)$ are the non-dimensional transverse displacement and lift force of the cylinder, respectively. The mass ratio is defined as $m^* = m/(m_f)$ where $m$ represents the total moving mass of cylinder and $m_f = \pi\rho D^2/4$ is mass of the fluid displaced by the cylinder. The reduced velocity $U^* = U/(f_n D)$ is based on the natural frequency of the cylinder in vacuum ($f_n = \frac{1}{2\pi}\sqrt{k/m}$). The damping ratio ($\zeta$) is defined as $\zeta = c/c_c$ where $c_c = 4\pi m f_n$ is the critical damping for the spring-mass system. The structure equation of motion is integrated in time using the following third order scheme

$$\dot{Y}^{n+1} = \dot{Y}^n + \frac{\Delta t}{12}\left[23\ddot{Y}^n - 16\ddot{Y}^{n-1} + 5\ddot{Y}^{n-2}\right], \qquad (3.4)$$

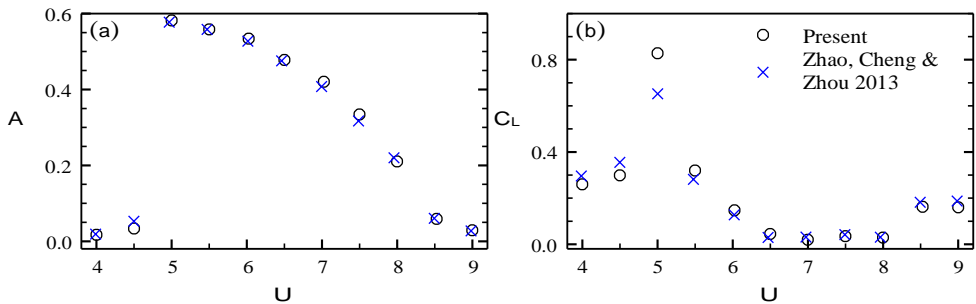

Figure 2: Comparison of (a) transverse vibration amplitude and (b) RMS lift coefficient of a circular cylinder with $m^* = 10$ having transverse and in-line degrees of freedom at $Re = 100$ with that of Zhao et al. (2013).

$$Y^{n+1} = Y^n + \frac{\Delta t}{12}\left[5\dot{Y}^{n+1} + 8\dot{Y}^n - \dot{Y}^{n-1}\right], \quad (3.5)$$

where $\Delta t$ represents the time step and the superscript $n$ is an integer which represents the current simulation time $t = n\Delta t$.

The coupled fluid and structure equations of motion are solved using a partitioned approach. In order to advance a simulation from time $n\Delta t$ to $(n + 1)\Delta t$, the structure equations are solved first using the current ($n$) and old ($n - k, k = 1, 2$) fluid forces and the cylinder is moved to its new ($n + 1$) location. The flow equation are then solved to obtain the new flow field. If the time step is sufficiently small then the solution obtained from the above process is very close to the actual solution of the coupled discrete equations. This scheme is known as explicit coupling. The explicit coupling is preferred due to its simplicity and less computational cost. However, the explicit coupling is stable for only large mass ratios ($m^* > 1$). An implicit coupling is used for smaller mass ratios. In the implicit coupling, the solution obtained by the explicit coupling is regarded as a first guess for the new solution. The procedure of the explicit coupling is applied to the first guess to obtain a second improved guess for the new solution. This iteration continues until the change between the two consecutive guesses becomes smaller than a predefined limit. In the present study, mass-ratio is taken as 10 and hence explicit coupling is used. For various channel heights, the cylinder motion is calculated for different combinations of the reduced velocity and damping ratio to find the optimal cases for power extraction.

### 3.3. Code validation

The sharp interface immersed boundary (SIIB) solver, used in the present work, had been validated previously for a number of benchmark problems. For example, Mittal et al. (2008) validated the flow solver for flow past a stationary circular cylinder, aerofoil and sphere. Bhardwaj & Mittal (2012) validated the SIIB based fluid-structure interaction (FSI) solver against the benchmark cylinder-flag problem proposed by Turek & Hron (2006). Soti et al. (2015) validated the SIIB solver for flow past stationary and moving heated circular cylinders. Garg et al. (2018) validated the present solver for vortex-induced vibrations (VIV) of circular cylinders transverse to the flow in isothermal and non-isothermal (to account for thermal buoyancy effects) flow conditions. Another validation study was performed in the present work on VIV of a circular cylinder having transverse and in-line degrees of freedom with respect to the free-stream flow direction. The mass-ratio of the cylinder is 10 and the Reynolds number is kept at 100. The natural frequencies of the cylinder for transverse and in-line motions are assumed equal. The computational domain is a rectangle with size $50D \times 40D$, where $D$ is the cylinder diameter. The center of the cylinder is located at $20D$ from the inlet and midway along the transverse direction.

| $\Delta X$ | 0.0400 | 0.0200 | 0.0100 |
|---|---|---|---|
| $A^*$ | 0.1957 | 0.2032 | 0.2042 |
| $\overline{P}$ | 0.8139 | 0.8765 | 0.8845 |

Table 1: Mesh independence study for $H = 2D$, $U^* = 1.82$ and $\zeta = 0.035$. Effect of grid size on vibration amplitude and average extracted power.

| $L_u/H$ | 1.0 | 2.0 | 5.0 |
|---|---|---|---|
| $A^*$ | 0.4975 | 0.4852 | 0.4854 |

Table 2: Domain independence study for $H = 4D$, $U^* = 4.0$ and $\zeta = 0.0$. Effect of distance of the cylinder from the channel inlet on vibration amplitude of the cylinder.

| $L_d/H$ | 8.0 | 13.0 | 15.0 |
|---|---|---|---|
| $A^*$ | 0.3702 | 0.3703 | 0.3696 |

Table 3: Domain independence study for $H = 2D$, $U^* = 1.67$ and $\zeta = 0.0$. Effect of distance of the cylinder from the channel outlet on vibration amplitude of the cylinder.

Figure 2 shows the transverse vibration amplitudes and root-mean-square (RMS) values of the lift coefficient at various reduced velocities. The present results are in good agreement with that of the Zhao *et al.* (2013).

### 3.4. Mesh and domain independence studies

The computational domain shown in figure 1 was meshed using a uniform grid of size $\Delta y$ in the *y*-direction. In the *x*-direction, a combination of a uniform grid of size $\Delta x = \Delta y$ which covers the region around the cylinder and a stretching non-uniform grid which covers the rest of the domain was used. The mesh independence study was performed by changing the mesh size $\Delta X = \Delta x/D$ to $0.04$, $0.02$ and $0.01$ in the uniform mesh region. Table 1 shows the vibration amplitude ($A^*$) of the cylinder and average extracted power ($\overline{P}$) for $H = 2D$, $U^* = 1.82$ and $\zeta = 0.035$ case using the three mesh sizes. Mesh size $\Delta X = 0.02$ was chosen for the simulations since refinement beyond 0.02 does not change the vibration amplitude by more than $1\%$.

A number of numerical simulation were done to find the distance of cylinder form the channel inlet ($L_u$) and channel length ($L_u + L_d$) and that did not effect the computational results significantly. It was found that $L_u > 2H$ and $L_d > 8H$ did not effect the vibration response of the cylinder significantly. Therefore, $L_u = 2H$ and $L_d = 8H$ were chosen for the present work. For example, table 2 shows effect of $L_u$ on the vibration amplitude of the cylinder for the case $H = 4D$, $U^* = 4.0$ and $\zeta = 0.0$. Similarly, table 3 shows the effect of $L_d$ on the vibration amplitude of the cylinder for the case $H = 2D$, $U^* = 1.67$ and $\zeta = 0.0$. The percentage change in the vibration amplitude is less than $1\%$ when $L_u$ and $L_d$ are increased beyond $2H$ and $8H$, respectively.

## 4. Results

The effect of blockage $b$ ($= D/H$) on vortex-induced vibration of the circular cylinder kept in a channel flow is investigated. The Reynolds number, based on inlet mean velocity $U$ and cylinder diameter $D$, and mass ratio are fixed at 150 and 10, respectively. A brief background on the vortex-induced vibrations (VIV) of a unconfined cylinder is presented in the next section which will be useful for the discussion on VIV of confined cylinder case.

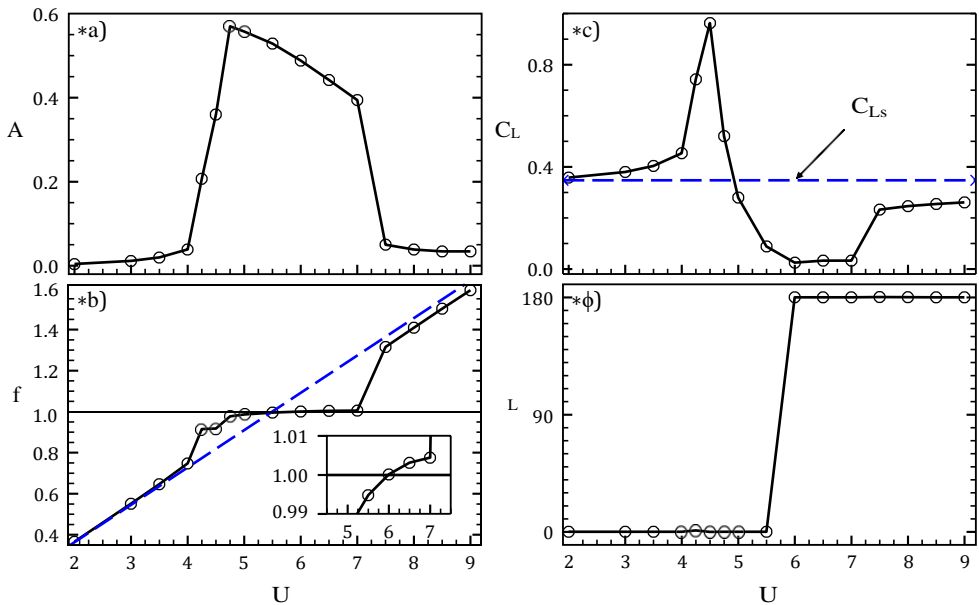

Figure 3: Vortex-induced vibration response of a unconfined circular cylinder of mass-ratio 10 at $Re = 150$. (a) Vibration amplitude, (b) normalized vibration frequency, (c) RMS lift force and (d) phase of the lift force with respect to the displacement.

### *4.1. VIV of a unconfined cylinder*

The vibration response of a circular cylinder kept in free-stream flow is plotted in figure 3 for $m^\star = 10$ and $Re = 150$. In the case of the unconfined cylinder, there exists a range of reduced velocity where the cylinder vibrates with a significant vibration amplitude (figure 3a). In this range of reduced velocity, the vortex shedding frequency remains close to the natural frequency of the cylinder (figure 3b where $f$ is vibration frequency normalized by the natural frequency of the cylinder). Therefore, this reduced velocity range ($4.0 < U^\star < 7.5$ in figure 3) is called the *lock-in region*. The cylinder vibrates in a near-sinusoidal fashion in the lock-in region. At low Reynolds number, the lock-in region is made of the initial and lower branch. The initial branch is the lower reduced velocity region ($4.0 < U^\star < 5.0$ in figure 3) where the cylinder motion is quasi-periodic. The interference of the vortex shedding frequency, corresponding to the stationary cylinder, and the natural frequency of the cylinder is responsible for the quasi-periodic nature. In the lower branch, the cylinder motion is dominated by only one forcing frequency as the vortex shedding frequency gets locked to the natural frequency of the cylinder. The cylinder motion is periodic in this branch. In the desynchronization region before the initial branch, the lift force asymptotically approaches the lift force on a stationary cylinder ($C_{Ls}$ in figure 3c) as the reduced velocity is decreased. The lift force increases rapidly with the reduced velocity in the initial branch and reaches its peak value which is much larger than $C_{Ls}$. In the lower branch, initially, the lift force decreases rapidly with the reduced velocity and attains a minimum value which is lower than $C_{Ls}$. Towards the end of the lower branch, the lift force slowly increases with the reduced velocity and settles at a value smaller than $C_{Ls}$ in the desynchronization region after the lower branch. Since the cylinder motion is periodic, the displacement and lift force on the cylinder can be expressed using Fourier series. In the lower branch, the cylinder displacement has a single frequency component.

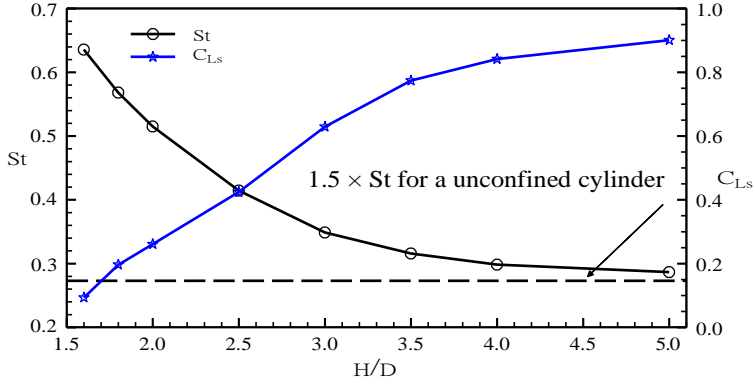

Figure 4: Strouhal number ($St$) and RMS lift coefficient ($C_{Ls}$) for a stationary circular cylinder of diameter ($D$) kept inside a channel of height $H$ at $Re = 150$.

If the cylinder displacement and lift force are assumed to be of the form $Y = A^* sin(2\pi f t/U^*)$ and $C_L = C_{L0} sin(2\pi f t/U^* + \phi_T)$, respectively, where $C_{L0}$ is the amplitude of the lift force and $\phi_T$ is its phase with respect to the displacement, and substituted into equation 3.3 then equating the cosine and sine terms gives the following equations

$$\frac{4\pi^3 m^* A^*}{(U^*)^2} f \zeta = C_{L0} sin(\phi_T), \tag{4.1}$$

$$\frac{2\pi^3 m^* A^*}{(U^*)^2} (1 - f^2) = C_{L0} cos(\phi_T). \tag{4.2}$$

For $\zeta = 0$, from equation 4.1, the phase difference $\phi_T$ is either $0°$ or $180°$ as shown in figure 3d. To find the phase of lift shown in figure 3d, fast Fourier transforms (FFT) of lift and displacement signals were calculated which provides the information on frequencies and phases of various sinusoids present in each signal. The phase difference was calculated by taking the difference between the phases of the two signals that corresponds to the fundamental frequency of displacement. From equation 4.2, the phase difference will be $0°$ for $f < 1$ and $180°$ for $f > 1$. This is evident from the zoom-in view of $f$ near $5.5 < U^* < 6.0$ shown in the inset of figure 3c.

### 4.2. Stationary confined cylinder

The flow around a stationary confined circular cylinder can provide insight on the possible effects of confinement on its VIV response. For this reason, this section will present flow characteristics for a stationary confined circular cylinder. The Strouhal number $St = f_v D/U$, where $f_v$ is the vortex shedding frequency, and lift coefficient ($C_{Ls}$) for the stationary cylinder is plotted in figure 4 as a function of the channel height. The dashed line in figure 4 represents Strouhal number for a stationary unconfined cylinder with $U_{max} = 1.5U$ as the reference velocity. The vortex shedding frequency curve asymptotes towards the dashed line with increasing $H$ due to reduction in the wall effect. As $H < \infty$, the wall effects are absent and the flow near the cylinder approaches the case of flow around a cylinder in the free stream of flow with the velocity $U_{max}$. The same is true for the lift force. As the channel height is decreased, the influence of the channel wall on the cylinder wake becomes stronger. The increase in wall proximity results in increased flow velocity near the cylinder due to increased blockage. The increasing local flow velocity is responsible for the increase in the vortex shedding frequency with a higher blockage, as seen in figure 4. On the other hand, wall proximity has a stabilizing effect on the wake behind the cylinder. Figure 5 shows the streamlines and vorticity contours of the flow for channel heights $H =$ **1.4**D, **2**D and **3**D. As seen in figure 5a, the vortex shedding is suppressed for $H =$ **1.4**D due to wall-wake interaction. Two symmetric re-circulation regions are formed on the downstream side of the cylinder. Two additional re-circulation regions are formed downstream of the cylinder at the channel walls with opposite sign vorticity as compared to their cylinder wake counterpart. As the channel height is increased (figure 5b and 5c), the channel wall is unable to stop the shear layers from becoming unstable and the vortex shedding starts. The stabilization effect of the wall on the cylinder wake can also be seen from the lift coefficient plotted in figure 4. The lift force decreases with decreasing channel height due to wake stabilization.

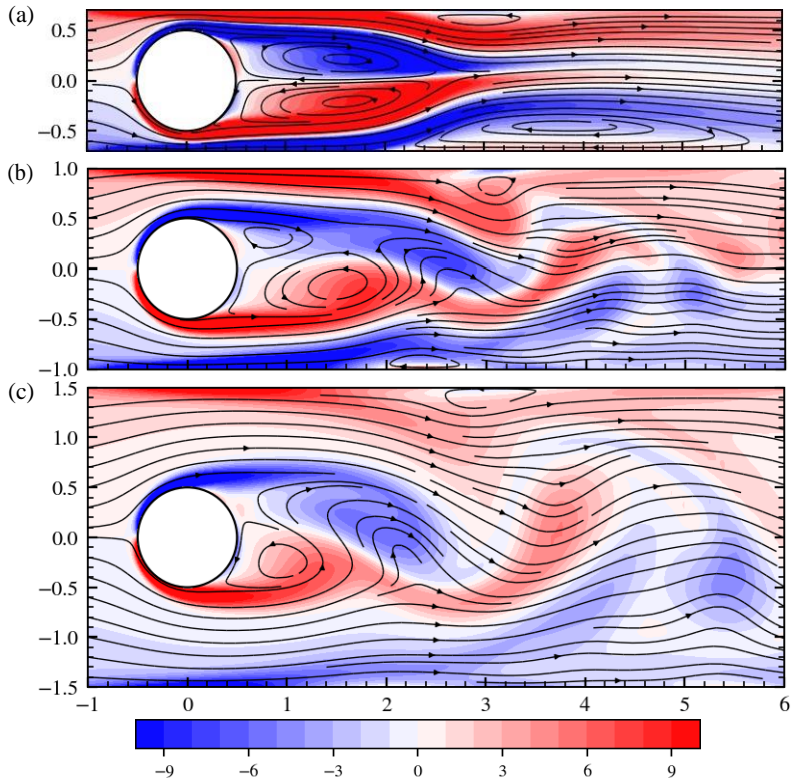

Figure 5: Vorticity contours and streamlines for the stationary circular cylinder at channel heights (a) **1.4**D, (b) **2**D and (c) **3**D at $Re = 150$. The vorticity scale is between -10 (blue) and 10 (red).

Based on observations for the stationary cylinder case, following predictions for VIV of the confined cylinder can be made. The lock-in region is known to occur near $U^{\star} =$ **1**/$St$ where $St$ is the Strouhal number for the stationary cylinder. Since the Strouhal number is found to increase with the increased blockage, the lock-in region is expected to shift towards the smaller values of reduced velocity with a decrease in the channel height. On the other hand, the decrease in the lift force with decreasing channel height may cause the vibration amplitude to decrease.

### 4.3. Undamped response

This section presents the effect of the channel blockage on the vibration response of the elastically mounted circular cylinder with zero damping. Figure 6a shows the non-dimensional vibration amplitude ($A^{\star}$) of the cylinder as a function of reduced velocity ($U^{\star}$) for three values channel heights $H =$ **2**D, **3**D and **4**D. The vibration amplitude was calculated using the following

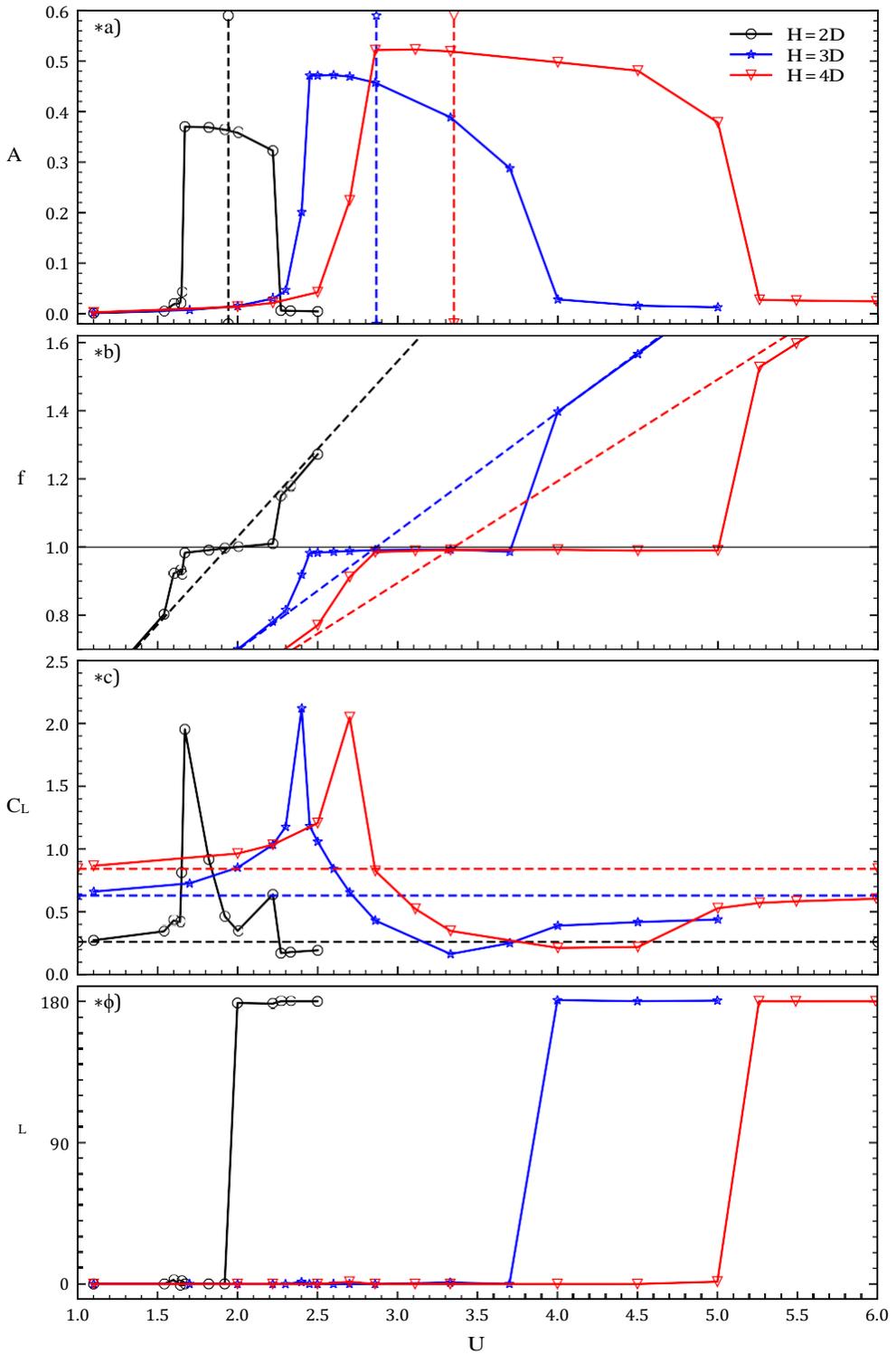

Figure 6: Vortex-induced vibration response of the circular cylinder of mass-ratio 10 for different blockages at *Re* = 150. (a) Vibration amplitude, (b) normalized vibration frequency, (c) RMS lift force and (d) mean phase of the lift force with respect to the displacement.

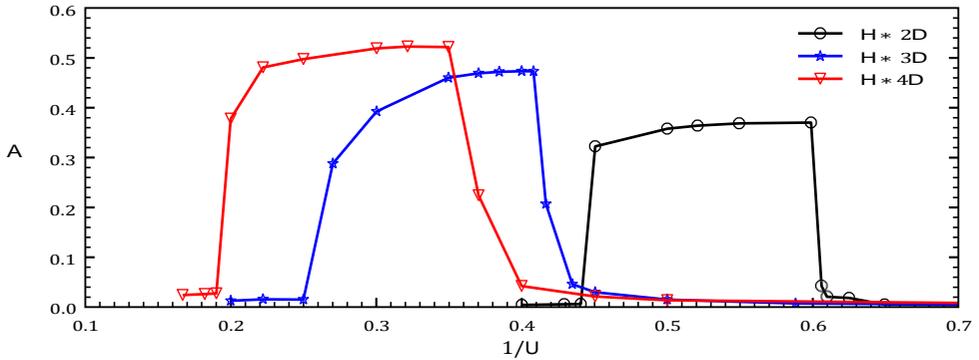

Figure 7: Vibration amplitude of the confined circular cylinder re-plotted against $1/U^*$.

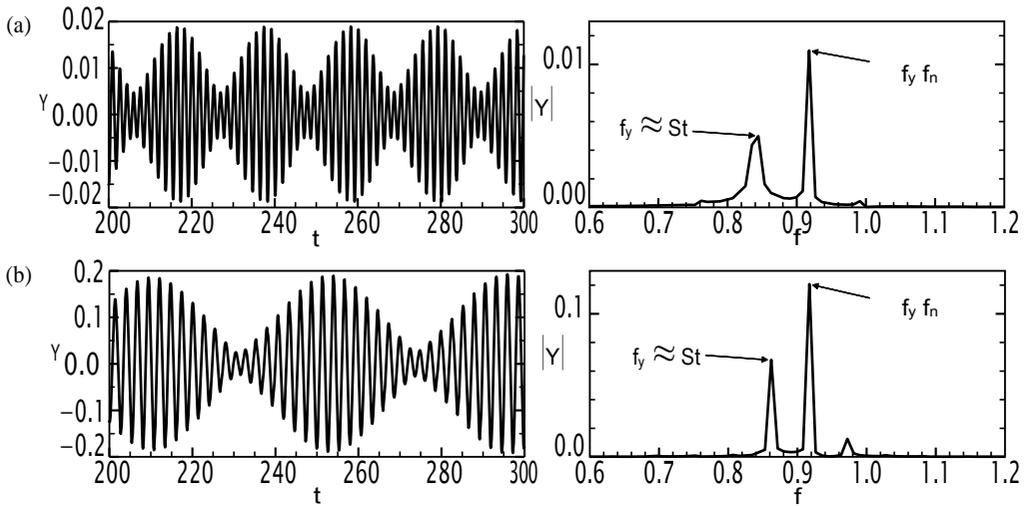

Figure 8: Cylinder displacement (left) in the initial branch and its Fourier transform (right) for (a) $H = 2D, U^* = 1.6$ and (b) $H = 3D, U^* = 2.4$. The frequency in the Fourier transform has been normalized by the natural frequency of the cylinder.

expression
$$A^* = \frac{1}{2} \sum (Y_{max} - Y_{min}) \qquad (4.3)$$
where $Y_{max}$ and $Y_{min}$ are the maximum and minimum values of the cylinder displacement after reaching a periodic/quasi-periodic state. The three channel cases show a similar trend for cylinder vibration. The cylinder vibrates with a small amplitude at lower $U^*$. As the reduced velocity is increased, the vibration amplitude of cylinder jumps to a higher value and continues to be at a significant fraction of the cylinder diameter for a range of $U^*$. Beyond a certain $U^*$, which depends on the channel height, the vibration amplitude again drops to a smaller value. The vertical broken lines in figure 6a represent $U^* = 1/St$ points for the three channel heights. As predicted, the vibration amplitude is found to decrease with the increased blockage. Figure 6b shows the vibration frequency of the cylinder normalized by its natural frequency. The dotted lines in figure 6b represent the vortex shedding frequency for the stationary cylinder case. The figure shows a lock-in region, similar to the unconfined case, for the three channel heights where the vortex shedding frequency (which is equal to the cylinder vibration frequency) is close to the natural frequency of the elastically mounted cylinder. Outside the lock-in region, the vortex shedding frequency follows the stationary case value. As expected, the lock-in region shifts towards the smaller values of reduced velocity due to the increased vortex shedding frequency for the stationary case as the channel height is decreased. The width of the lock-in region appears to decrease with the channel height. For further investigation, figure 6a is reproduced in figure 7 with the x-axis changed to $1/U^*$. From figure 7, it is clear that the apparent change in width of the lock-in region is due to the choice of the independent variable against which the vibration amplitude is presented. On the frequency axis, the width of the lock-in region is similar for all three channel heights.

Figure 6c shows the root mean square (RMS) of the non-dimensional lift force ($C_L$) acting on the cylinder. The horizontal broken lines in the figure represent the lift force on a stationary cylinder ($C_{Ls}$). The behavior of the lift force on the confined cylinder is also similar to that of the unconfined case. In all three channel cases, the lift force is equal to that for a stationary cylinder ($C_{Ls}$) at small reduced velocities. With an increase in the reduced velocity, the lift force first rapidly increases to a high value and then decreases to a value smaller than $C_{Ls}$. Towards the end of the synchronization region, the lift force increases slowly with the reduced velocity and approaches a value smaller than $C_{Ls}$ in the desynchronization region. The phase of the lift force was also investigated in the previous studies on VIV of an unconfined cylinder. The mean phase of the lift force with respect to the cylinder displacement is plotted in figure 6d. The lift force is in phase with the displacement in the synchronization region where the cylinder vibrates with significant amplitude. The phase of lift force jumps to $180°$ at the end of the synchronization region.

Due to the similarities discussed in the previous paragraphs, the branching of the vibration response of the confined cylinder is also similar to that of the unconfined case. Presence of the initial and the lower branches in the vibration response of the confined cylinder can be claimed based on following evidence. Figure 8 shows the cylinder displacement as a function of time (on left) and its Fourier transform (on right) for channel height $H = 2D$ and $H = 3D$. The vibration response in both cases is quasi-periodic. Two significant peaks can be seen in the Fourier transform: one corresponding to the natural frequency of the cylinder ($f = 0.92$ due to the added-mass effect) and the other one due to the vortex shedding frequency for a stationary cylinder. These are the characteristics of the initial branch. Note that as the channel height is decreased, the jump in the vibration amplitude at the start of the initial branch is steeper and the width of the initial branch has decreased. The rapid increase in the lift force in figure 6c at small reduced velocity also supports the presence of the initial branch. The lower branch starts when the lift force is maximum. The displacement of the cylinder is periodic in this region and only a single frequency is present in its spectrum (not shown here). The vibration amplitude of the cylinder is also larger in this branch as compared to that in the initial branch. The vibration frequency of the cylinder remains constant in the lower branch.

Figure 9 shows vorticity contours of the flow in channels of heights $2D$ and $3D$ when the cylinder has maximum vibration amplitude. The flow is shown at two instances when the cylinder is at its mean position ($Y = 0$) and moving up ($\dot{Y} > 0$) or down ($\dot{Y} < 0$). The vortex shedding pattern in both the cases is 2S where two counter-rotating vortices are shed per cycle of oscillation of the cylinder.

### 4.4. Damped response

The vibrating cylinder has some kinetic energy, a part of which can be extracted by using a power generator. The energy extraction process can be considered as an energy dissipater connected to the cylinder. Soti *et al.* (2017) have shown that a linear electrical generator can be modelled as a damper with spatially varying damping ratio. It was also shown that a much simpler power extraction model of a damper with constant damping ratio could correctly predict the average values but not the temporal behaviour of the extracted power. The present work is focused on the impact of confinement on the maximum average power extracted by the cylinder. Therefore, the damper with a constant damping ration model is chosen for modelling the power extraction process in the present work for the sake of simplicity. The power extraction calculations were performed by choosing a set of damping ratio values and changing the reduced velocity over the synchronization region obtained from the undamped response for each member of that set.

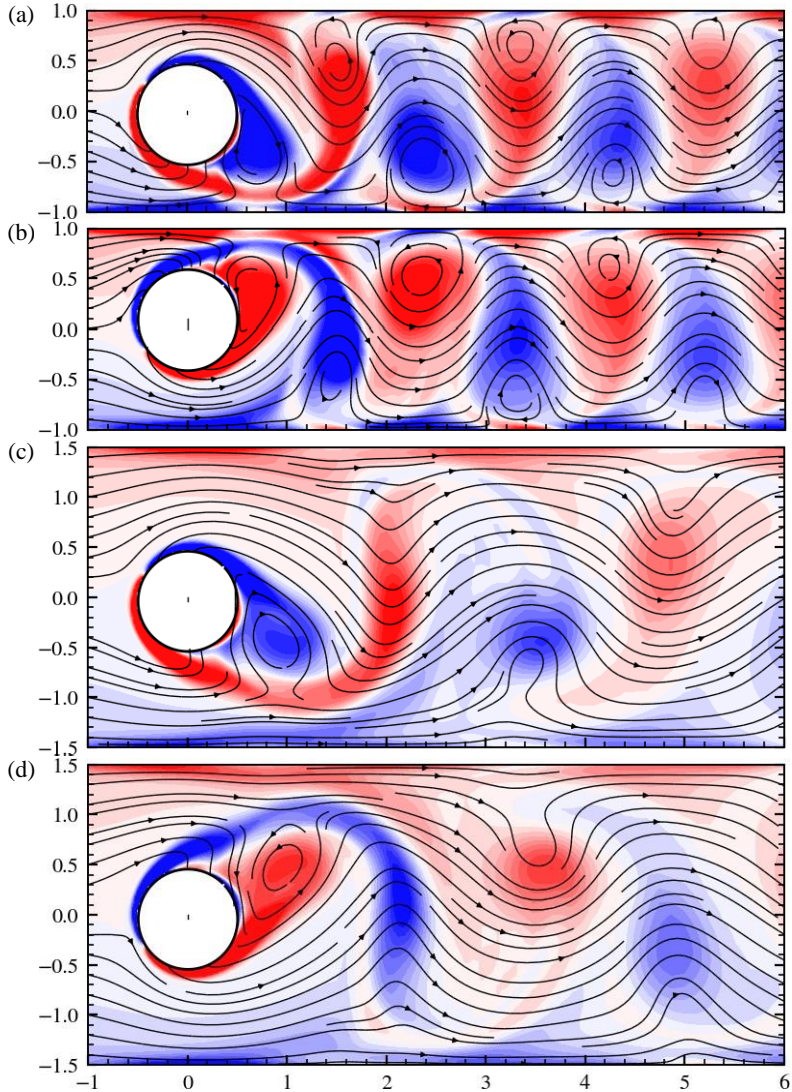

Figure 9: Vorticity contours and streamlines for the moving circular cylinder at $Re = 150$. (a & b) $H = 2D$ at $U^* = 1.67$ and (c & d) $H = 3D$ at $U^* = 2.60$. The cylinder is at its mean position while moving up (a & c) and moving down (b & d). The vorticity scale is between -10 (blue) and 10 (red).

The effect of damping on the vibration response of the confined cylinder is plotted in figure 10 for three channel heights $H = 2D, 2.5D$ and $3.5D$. As non-zero damping is introduced into the system, some part of the kinetic energy of the cylinder is dissipated by the damper. Therefore, at any reduced velocity in the synchronization region, the vibration amplitude of the cylinder decreases monotonically with increasing damping as seen in figure 10a, 10f and 10k. Also notice that, at any value of damping, there is a critical reduced velocity at which the vibration amplitude is maximum. It will be referred to as the peak vibration amplitude ($A^*$) at the corresponding damping where $\alpha = m^*\zeta$ is the mass-damping parameter. The peak vibration amplitude is important from the structural design point of view so that a structure can be designed to withstand the worst-case scenario of large scale vibrations. Figure 11a shows the peak vibration amplitude as a function of the mass-damping parameter for channel heights $H =$ **2**$D$, **2.5**$D$, **3**$D$ and **4**$D$. Evidentially, there is a monotonic decay of the peak vibration amplitude with increasing damping for all channel cases. The vibration frequency in the synchronization region decreases with increasing damping (see figure 10b, 10g and 10l) since the natural frequency of the spring-mass system decrease with damping.

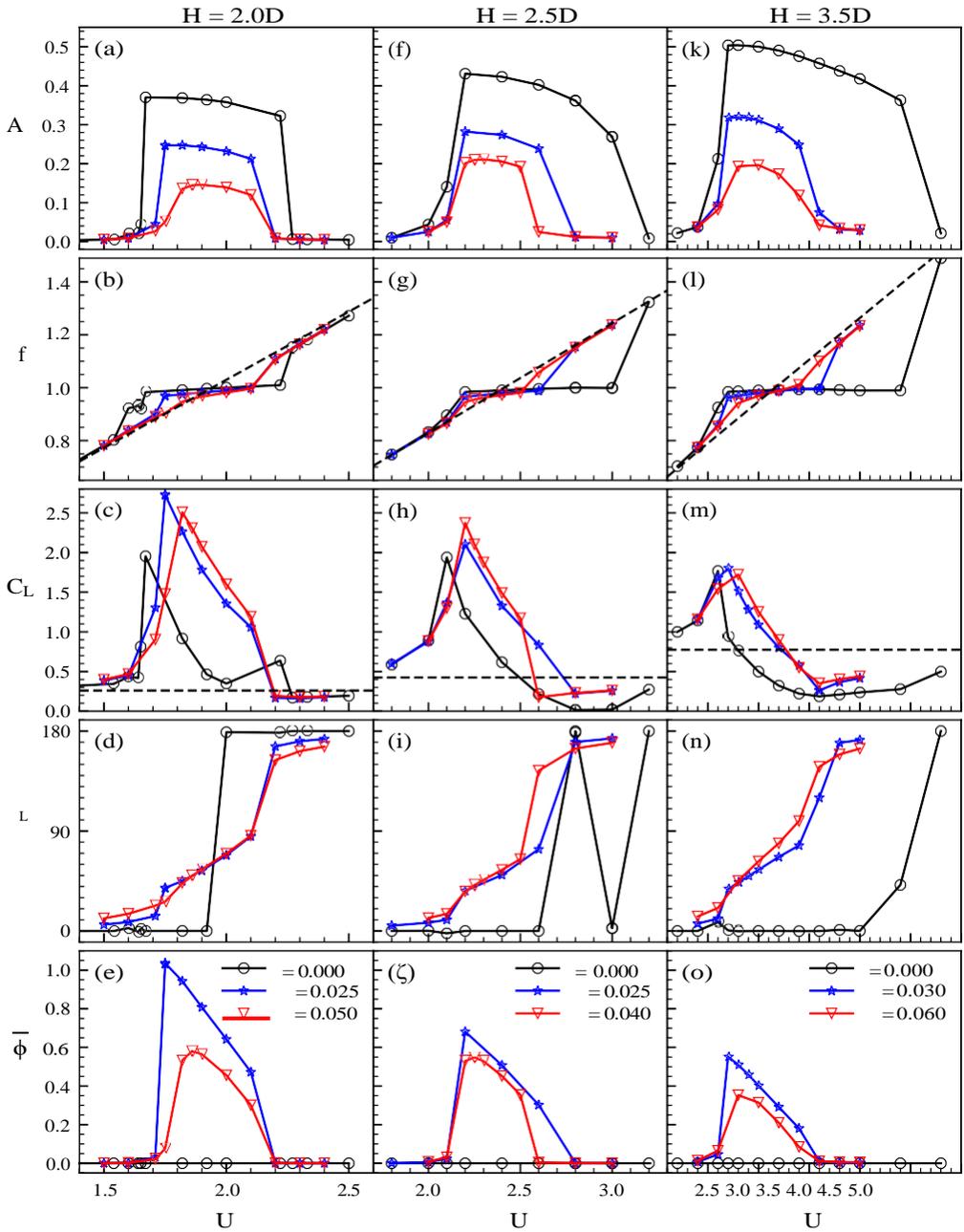

Figure 10: Effect of damping on the VIV response of the confined cylinder: (a, f k) Vibration amplitude, (b, g, l) normalized vibration frequency, (c, h, m) RMS lift force, (d, l, n) mean phase of the lift force with respect to the displacement and (e, j, o) average extracted power.

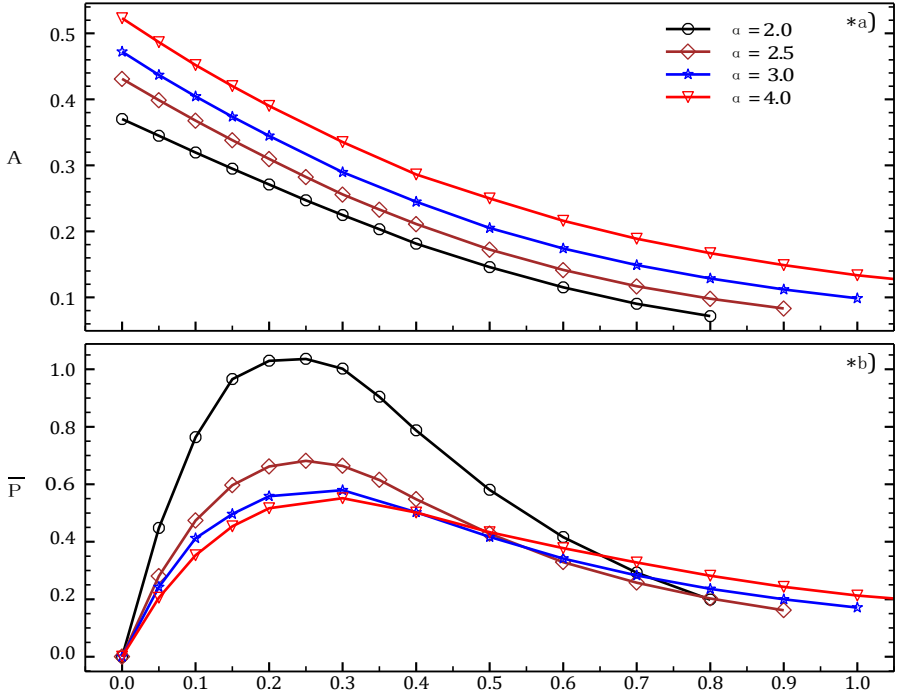

Figure 11: (a) Peak vibration amplitude, (b) peak average extracted power and (c) power extraction efficiency for the circular cylinder confined in the channels of heights between **2**D to **4**D.

### 4.5. Power extraction

The extracted power is equal to the power dissipated by the damper which can be calculated by taking the product of the damping coefficient (c) and square of the cylinder velocity ($\dot{y}$). The non-dimensional expression of the dissipated power is

$$P(t) = \frac{c\dot{y}^2}{\frac{1}{2}\rho U^3 D}, \qquad (4.4)$$

which can also be written in terms of the non-dimensional parameters as

$$P(t) = \frac{2\pi^2 m^* \zeta}{U^*} \dot{Y}^2 \qquad (4.5)$$

Equation 4.5 gives the instantaneous power extracted by the cylinder. Since the cylinder undergoes a periodic motion, the average extracted power can be defined as

$$\bar{P}(\alpha, U^*) = \frac{2\pi^2 \alpha}{U^*} \dot{Y}^2_{rms}, \qquad (4.6)$$

where $\alpha = m^*\zeta$ is the mass-damping parameter and the root-mean-square (RMS) velocity of the cylinder ($Y_{rms}$) is defined as

$$Y_{rms} = \sqrt{\frac{1}{T}\int_0^T \dot{Y}^2 dt}, \quad (4.7)$$

where $T$ represented the time period of oscillation. In the present numerical work, the average was taken over several cycles of oscillations to reduce the time discretization error. Notice that the RMS velocity of the cylinder is a function of $\alpha$ and $U^*$. If the cylinder displacement is approximated the function $Y = A^* \sin(2\pi f t / U^*)$ then the average extracted power is given by the following expression

$$\bar{P}(\alpha, U^*) = 4\pi^4 \frac{(fA^*)^2}{(U^*)^3}\alpha \quad (4.8)$$

Figure 10e, 10j and 10o shows the average extracted power as a function of the reduced velocity for three values of damping ratio at channel heights $H = 2D, 2.5D$ and $3.5D$, respectively. The extracted power is negligible in the desynchronization region due to small vibration amplitude. There is a jump in the extracted power near the start of the lower branch, which corresponds to the jump in the vibration amplitude. The vibration amplitude and the extracted power are decreasing functions of the reduced velocity. Therefore, the extracted power reduced rapidly with the reduced velocity. At any value of the mass-damping parameter, there is an optimal reduced velocity at which the average extracted power is maximum. It will be referred to as the peak average extracted power ($\bar{P}_\alpha$) at the corresponding mass-damping. Figure 11b shows the peak average extracted power as a function of the mass-damping parameter for channel heights $H = 2D, 2.5D, 3D$ and $4D$. From equation 4.8, the extracted power is zero at zero mass-damping. It will also be zero at a very large value of mass-damping ($\alpha < \infty$) since the vibration amplitude will approach zero. Therefore, as seen in figure 11b, the extracted power is maximum at some optimal intermediate value of mass-damping. This fact is true for all channel cases. This maximum value of the average extracted power will be referred to as the maximum extracted power ($\bar{P}_m$) for that channel height. Figure 12 shows the maximum vibration amplitude $A_0^* (= A_{\alpha=0}^*)$ and maximum extracted power $\bar{P}_m$ for different channel heights. The error bars on maximum extracted power represent **±20%** variations.

Assuming that the peak vibration amplitude shown in figure 11a can be expressed as

$$A_\alpha^* = A_0^* e^{-c_1 \alpha}, \quad (4.9)$$

where $c_1$ is a constant which can be determined by performing the least square fit on the data in figure 11a for each channel height. Then the peak average extracted power can be expressed as

$$\bar{P}_\alpha = 4\pi^4 \frac{fA_0^{*2}}{U_{c\alpha}^{*3}} \alpha e^{-2c_1 \alpha}, \quad (4.10)$$

where $U_{c\alpha}^*$ is the critical reduced velocity at which the average extracted power is maximum for mass-damping $\alpha$. The peak average extracted power is maximized when the derivative $\partial \bar{P}_\alpha / \partial \alpha$ is equal to zero. This gives

$$\bar{P}_m = \frac{4\pi^4 fA_0^{*2}}{e \cdot U_{c\alpha}^{*3}} \alpha_c, \quad (4.11)$$

where $\alpha_c = 1/2c_1$ is the optimal mass-damping parameter. Soti *et al.* (2017) using numerical simulations found $\alpha_c = 0.28$ at $Re = 150$ for an unconfined circular cylinder. Soti *et al.* (2018) using water channel experiments found $\alpha_c$ in the range of 0.2 to 0.3 for Reynolds number ranging between 1500 to 6000. In figure 11b, the optimal mass-damping parameter can be seen to range between 0.25 to 0.3. The critical vibration amplitude at the optimal mass-damping is $A^* = A_0^*/\sqrt{e}$.

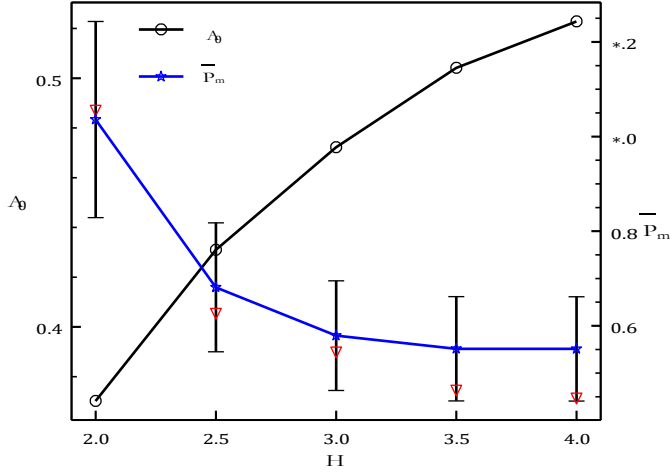

Figure 12: The maximum vibration amplitude $A_0^*$ and maximum extracted power $\overline{P}_m$ for different channel heights. The errorbars represent **±20%** variation in $P_m$. The inverted triangles represent the values of $P_m$ calculated using equation 4.12 with $\alpha_c =$ **0.25**.

The maximum power is extracted in the lock-in region. Therefore, the frequency ratio can be approximated as $\hat{f} = 1$ in equation 4.11. The maximum vibration amplitude $A_0^*$ is obtained from the undamped response. For simplicity, the critical reduced velocity $U_c^*$ is approximated by the reduced velocity $U^*$ at which maximum vibration amplitude $A_0^*$ is obtained. This can also be obtained from the undamped response. Using these approximations, the expression for the maximum extracted power simplifies to

$$\overline{P}_m = \frac{4\pi^4}{e} \cdot \frac{A_0^{*2}}{U_{c0}^{*3}} \alpha_c. \tag{4.12}$$

In the above equation, every variable except the optimal mass-damping can be determined from the undamped response of the cylinder. In figure 12, the maximum extracted power predicted by equation 4.12 with $\alpha_c =$ **0.25** is shown by the inverted triangles. The predicted values are within **±20%** range of the values obtained from the simulations. For $H =$ **2***D*, the error in the predicted value is small since the actual $\alpha_c$ is 0.25 while the error is large for $H =$ **4***D* since the actual value of $\alpha_c$ is 0.3 which is **20%** more than the 0.25. Therefore, the optimal mass-damping is the most significant source of error in equation 4.12. From low *Re* simulations of Soti *et al.* (2017), $A^* =$ **0.59**, $A_0^* =$ **0.36**, $U^* =$ **4.8**, $\alpha_c =$ **0.28** and $\overline{P}_m =$ **0.128** at *Re* = 150. From high *Re* experiments of Soti *et al.* (2018), $A^* =$ **0.87**, $A_0^* =$ **0.62**, $U^* =$ **5.0** (5.8 with added mass), $\alpha_c =$ **0.22** and $\overline{P}_m =$ **0.184** for $Re =$ 2220 – 6661 (set 2). Using $\alpha_c =$ **0.25**, the values of maximum extracted power predicted by equation 4.12 are 0.113 and 0.217 for low and high Reynolds number cases, respectively. These predictions differ by **12%** and **18%** from the actual values for low and high Reynolds number cases, respectively. Therefore, equation 4.12 provides an accurate means of predicting the maximum power extracted by the cylinder using its undamped vortex-induced vibration response.

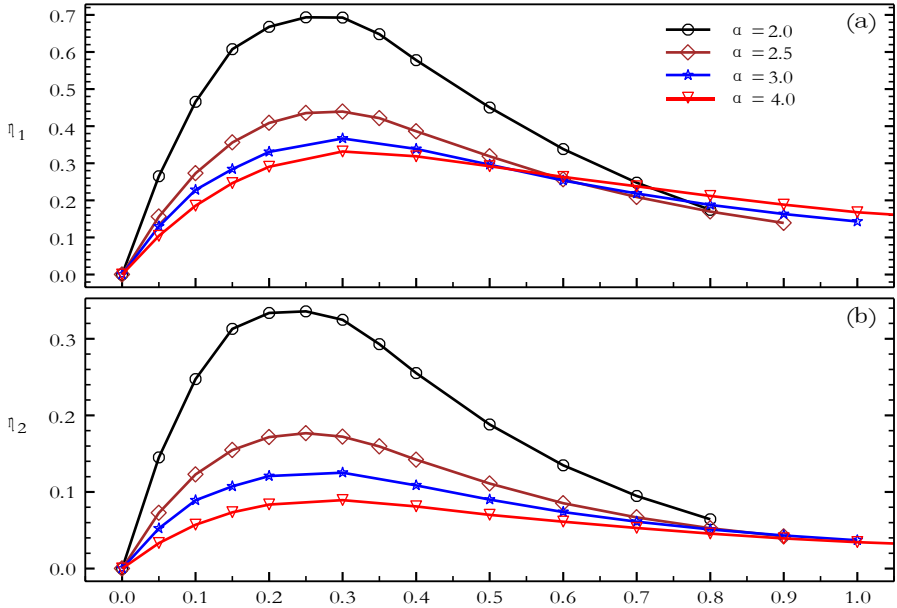

Figure 13: power extraction efficiency for the circular cylinder confined in the channels of heights between **2**$D$ to **4**$D$.

### 4.6. *Power extraction efficiency*

The power extraction efficiency ($\eta_a$) is defined as the ratio of extracted power to the maximum power that can be extracted from the flow. Maximum energy is extracted when the flow is brought to complete rest and all its kinetic energy is harnessed. In a uniform flow, $qU_\infty$ mass of fluid will cross a unit area per second with velocity $U_\infty$ where $q$ is the fluid velocity. Therefore, $qU^3/2$ is the maximum flow power per unit area that can be extracted. For an unconfined transversely vibrating cylinder, the maximum extractable power is equal to $(D+2A)qU^3/2$ (assuming the unit length of the cylinder) where $A$ is its vibration amplitude. Therefore, the mathematical expression of power extraction efficiency is

$$\eta_1 = \frac{\overline{P}}{1 + 2A\bar{x}} \quad (4.13)$$

The efficiency is a function of the reduced velocity and mass-damping and has a peak value for any mass-damping at an optimal reduced velocity. In general, the optimal mass-damping for the maximum efficiency is greater than or equal to the optimal mass-damping for maximum power extraction. Soti *et al.* (2017) and Soti *et al.* (2018) found maximum power extraction efficiency ($\eta_1$) to be **7.44%** and **8.21%** for low and high *Re* cases, respectively.

The aforementioned definition of power extraction efficiency can be extended to the confined cylinder case by replacing the free stream velocity ($U_\infty$) with the mean flow velocity ($U$) of the channel. Note that this choice keeps the volume flow rate the same in confined and unconfined cases. With that consideration, equation 4.13 can also be used to calculate the power extraction efficiency of the confined circular cylinder. The efficiency ($\eta_1$) in this case represents the flow power extracted by the confined cylinder as a fraction of the maximum flow power that could be harnessed by an unconfined cylinder vibrating with the same vibration amplitude. Figure 13a shows the peak power extraction efficiency ($\eta_1$) for the confined cylinder inside channel heights $H$ = **2**$D$, **2.5**$D$, **3**$D$ and **4**$D$. There is a rapid increase in efficiency as the channel height is reduced.

The maximum power extraction efficiency ($\eta_1$) for channel height $H = 2D$ is **69.33%**, which is close to an order of magnitude higher than that for an unconfined cylinder.

One could also consider the total flow energy available inside the channel to define the power extraction efficiency of the cylinder. The maximum extractable flow power, in that case, is equal to $27HqU^3/35$. Therefore, the second power extraction efficiency ($\eta_2$) is defined as

$$\eta_2 = \frac{35\,\bar{P}}{54\,H/D} \tag{4.14}$$

Figure 13b shows the peak power extraction efficiency ($\eta_2$) for the confined cylinder inside channel heights $H = 2D, 2.5D, 3D$ and $4D$. The second efficiency also increases rapidly with the reduction in the channel height. The maximum value of the second efficiency for channel height $H = 2D$ is **33.57%** implying that close to **34%** flow energy of the channel is harnessed by the cylinder.

## 5. Conclusions

Vortex-induced vibration (VIV) of a circular cylinder kept inside a parallel plate channel was investigated numerically at Reynolds number 150. The cylinder was elastically mounted with a linear spring along the direction transverse to the flow. The mass ratio of the cylinder was fixed at 10 and the channel height was varied from $2D$ to $4D$ with an increment of $0.5D$. The reduced velocity was varied over a range to determine the lock-in region. The objective of the study was to evaluate the effect of confinement on the vibration response and flow power extraction capability of the cylinder. The power extraction process was modeled as a damper with spatially constant damping ratio attached to the cylinder.

The vibration amplitude of the cylinder is found to decrease with the channel height. On the reduced velocity ($U^*$) axis, the range of the lock-in region is found to decrease with decreasing channel height. However, on the reduced frequency ($f^* = 1/U^*$) axis, the width of the lock-in region is the same for all channel heights considered. For the stationary confined cylinder, the flow velocity in the proximity of the cylinder increases with reducing channel height due to the increased blockage which causes the vortex shedding frequency ($St$) to increase with decreasing channel height. Since the lock-in happens close to the reduced velocity of $1/St$, the lock-in region shifts towards smaller reduced velocity as the channel height is reduced.

The vibration amplitude decreases monotonically with the damping for all channel heights. The extracted power is a function of the reduced velocity and mass-damping ($m^*\zeta$). There is an optimal combination of the two parameters at which the extracted power is maximized. The maximum extracted power is found to increase rapidly with decreasing channel height. For channel height $H = 2D$, the maximum flow power extracted by the cylinder is an order of magnitude larger as compared to what it would extract in an open domain with free stream velocity equal to the channel mean velocity. The percentage of channel flow energy harnessed by the cylinder increases rapidly with the decreasing channel height. The maximum extracted power is **34%** of the total flow energy available in the channel for channel height $2D$, which is 3.8 times larger than the same for channel height $4D$.

By assuming that the peak vibrating amplitude decreases exponentially with the mass-damping, an expression to predict the maximum extracted power from the undamped response of a confined/unconfined cylinder is derived. By taking the optimal mass-damping to be equal to 0.25, the derived expression is able to predict the maximum power extraction within **±20%** of the actual values obtained from present and previous numerical and experimental studies.

## 6. Acknowledgements

Sir, please add the funding agencies that you want to acknowledge.